\documentclass[twoside,10pt,letterpaper]{article}
\usepackage{amsmath,chemarrow}
\usepackage{graphicx}

\newcommand{\bea}{\begin{eqnarray}}

\newcommand{\eea}{\end{eqnarray}}

\begin{document}
%

\title{Astumian's paradox revisited}
\author{R. Dean Astumian\\ Department of Physics and Astronomy\\ University of
Maine\\ Orono, Maine 04469-5709\\ e-mail:
astumian@maine.edu\vspace{1ex}l}
\date{}

\maketitle

\markboth{The Astumian's Paradox}{Piotrowski and S\l adkowski}

\pagestyle{myheadings}


\begin{abstract}
I give a simple analysis of the game that I previously published in Scientific American which shows the
paradoxical behavior whereby two losing games randomly
combine to form a winning game. The game, modeled on a random walk, requires only two states and is described by a first-order Markov process.\end{abstract}
 \vspace{5mm}

\noindent In 1986--7 my colleagues and I posed the
following ``paradox."\cite{west_pnas86,ast_pnas87} Consider a
random walk on a cycle of bases governed by one of two sets of
transition constants for stepping between the bases. Using either
set alone, the walk is biased to favor completion of a
counterclockwise cycle. However, either periodic\cite{west_pnas86}
or random\cite{ast_pnas87} alternation between the two sets causes
net clockwise cycling! 

To illustrate as simply as possible this paradoxical behavior, I
published \cite{ast_sciam01} a very simple game played with a
checker stepping on part of a checkerboard. The stepping is decided
by the roll of a pair of dice. Alternation between two sets of
rules for the stepping is achieved by flipping a coin. In a recent
paper\cite{piot_lanl04} Piotrowski and Sladowski discussed this
``Astumian paradox" and asserted that my analysis was flawed, and
that the game does not display the paradoxical behavior that I
claimed. Here I show that the claims Piotrowski and Sladowski are wrong and that my game does show the paradoxical behavior first demonstrated in
Refs.1 and 2.

The game I published in Scientific American consists of stochastic jumping between five different states 1, ..., 5 and is based on the diagram below,

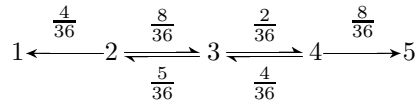
\begin{displaymath}
1 \autoleftarrow{$\frac{4}{36}$}{} 2 \autorightleftharpoons{$\frac{8}{36}$}{$\frac{5}{36}$}3\autorightleftharpoons{$\frac{2}{36}$}{$\frac{4}{36}$} 4 \autorightarrow{$\frac{8}{36}$}{}5
\end{displaymath}

where the numbers written above and below the arrows are the probabilities of transitions between neighboring states.  The player wins if he/she winds up at state five, and loses if he/she reaches state 1.  If the player  starts from state 3 the ratio of the probability of losing to the probability of winning is $5*4/(8*2) = 20/16 = 5/4$ as reported by Piotrowski and Sladowski as well as by me \cite{ast_sciam01}.  However, in their diagram (1), $P\&S$ made a critical error - they implicitly and without justification normalized the probabilities such that $p(i \rightarrow i+1) + p(i \rightarrow i-1) = 1, \,\,\,\,\, i=2,3,4$.  Not only is this explicitly different than the case in the game I published (diagram (1) above) but is totally and absolutely incorrect for the physical system - a random walk on a biochemical network - that the game was designed to illustrate!  Indeed, in a monte carlo simulation of a kinetic network the transition probabilities out of a state should be much less than one $p(i \rightarrow i+1) + p(i \rightarrow i-1) << 1$ so that in any iteration at most one, and most often zero, transitions occur.

The same numerical result for the ratio of losses to wins is obtained for the game in diagram (2)

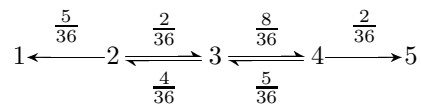
\begin{displaymath}
1 \autoleftarrow{$\frac{5}{36}$}{} 2 \autorightleftharpoons{$\frac{2}{36}$}{$\frac{4}{36}$}3\autorightleftharpoons{$\frac{8}{36}$}{$\frac{5}{36}$} 4 \autorightarrow{$\frac{2}{36}$}{}5
\end{displaymath}

Both games (1) and (2) result in more losses than wins.  A modified game, where the probabilities are the arithmetic mean of the probabilities for the above two games, is shown below:

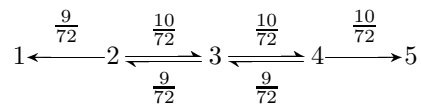
\begin{displaymath}
1 \autoleftarrow{$\frac{9}{72}$}{} 2 \autorightleftharpoons{$\frac{10}{72}$}{$\frac{9}{72}$}3\autorightleftharpoons{$\frac{10}{72}$}{$\frac{9}{72}$} 4 \autorightarrow{$\frac{10}{72}$}{}5
\end{displaymath}

Even a superficial inspection of the above diagram shows that my analysis published in Scientific American is correct, and the criticism of Piotrowski and Sladowski is wrong: The transition from states 2,3, and 4 to the right are more probable than to the left resulting in more wins than losses.  It seems that an inability to accurately transcribe the probabilities published in my Scientific American and an inability to do simple probability calculations for a pair of dice have resulted in drawing wrong conclusions by Profs. Piotrowski and Sladowski.

\end{document}